# FINGERPRINT RECOGNITION USING MINUTIA SCORE MATCHING

RAVI. J[*], K. B. RAJA[**], VENUGOPAL. K. R[***]

*Dept. of Electronics and Communication Engg, Global Academy of Technology, Bangalore – 98

**Dept. of Electronics and Communication Engineering,

University Visvesvaraya College of Engineering, Bangalore-560001

***Dept. of Computer Science and Engineering,

University Visvesvaraya College of Engineering, Bangalore-560001

**ABSTRACT:** The popular Biometric used to authenticate a person is Fingerprint which is unique and permanent throughout a person's life. A minutia matching is widely used for fingerprint recognition and can be classified as ridge ending and ridge bifurcation. In this paper we projected Fingerprint Recognition using Minutia Score Matching method (FRMSM). For Fingerprint thinning, the Block Filter is used, which scans the image at the boundary to preserves the quality of the image and extract the minutiae from the thinned image. The false matching ratio is better compared to the existing algorithm.

## I. Introduction

Biometric systems operate on behavioral and physiological biometric data to identify a person. The behavioral biometric parameters are signature, gait, speech and keystroke, these parameters change with age and environment. However physiological characteristics such as face, fingerprint, palm print and iris remains unchanged through out the life time of a person. The biometric system operates as verification mode or identification mode depending on the requirement of an application. The verification mode validates a person's identity by comparing captured biometric data with ready made template. The identification mode recognizes a person's identity by performing matches against multiple fingerprint biometric templates. Fingerprints are widely used in daily life for more than 100 years due to its *feasibility, distinctiveness, permanence, accuracy, reliability, and acceptability*. Fingerprint is a pattern of ridges, furrows and minutiae, which are extracted using inked impression on a paper or sensors. A good quality fingerprint contains 25 to 80 minutiae depending on sensor resolution and finger placement on the sensor. The false minutiae are the false ridge breaks due to insufficient amount of ink and cross-connections due to over inking. It is difficult to extract reliably minutia from poor quality fingerprint impressions arising from very dry fingers and fingers mutilated by scars, scratches due to accidents, injuries. Minutia based fingerprint recognition consists of *Thinning, Minutiae extraction, Minutiae matching and Computing matching score.*

Motivation: The motivation behind the work is growing need to identify a person for security. The fingerprint is one of the popular biometric methods used to authenticate human being. The proposed fingerprint verification FRMSM provides reliable and better performance than the existing technique.

Contribution: In this paper we used Fingerprint Recognition using Minutia Score Matching method with the help of MATLAB codes. Minutiae are extracted from the thinned image for both template and input image. Finally both the images are subjected to matching process and matching score is computed.

Organization: This paper is organized into the following sections. Section II is an overview of the related work, in section III describes Model for fingerprint recognition in detail. Section IV gives the algorithm. In section V performance analysis and results are discussed and finally in section VI give the conclusions.

## II. RELATED WORK

G. Sambasiva Rao et al., [1] proposed fingerprint identification technique using a gray level watershed method to find out the ridges present on a fingerprint image by directly scanned fingerprints or inked impression. Robert Hastings [2] developed a method for enhancing the ridge pattern by using a process of oriented diffusion by adaptation of anisotropic diffusion to smooth the image in the direction parallel to the ridge flow. The image intensity varies smoothly as one traverse along the ridges or valleys by removing most of the small irregularities and breaks but with the identity of the individual ridges and valleys preserved. Jinwei Gu, et al., [3] proposed a method for fingerprint verification which includes both minutiae and model based orientation field is used. It gives robust discriminatory information other than minutiae points. Fingerprint matching is done by combining the decisions of the matchers based on the orientation field and minutiae.

[*] ravignkere@rediffmail.com
Phone: 09481252624
Fax: 080 – 28603157
[**] raja_kb@yahoo.com
Phone: 09448062034

V. Vijaya Kumari and N. Suriyanarayanan [4] proposed a method for performance measure of local operators in fingerprint by detecting the edges of fingerprint images using five local operators namely Sobel, Roberts, Prewitt, Canny and LoG. The edge detected image is further segmented to extract individual segments from the image. Raju Sonavane, and B.S. Sawant [5] presented a method by introducing a special domain fingerprint enhancement method which decomposes the fingerprint image into a set of filtered images then orientation field is estimated. A quality mask distinguishes the recoverable and unrecoverable corrupted regions in the input image are generated. Using the estimated orientation field, the input fingerprint image is adaptively enhanced in the recoverable regions.

Eric P. Kukula, et al., [6] purposed a method to investigate the effect of five different force levels on fingerprint matching performance, image quality scores, and minutiae count between optical and capacitance fingerprint sensors. Three images were collected from the right index fingers of 75 participants for each sensing technology. Descriptive statistics, analysis of variance, and Kruskal-Wallis nonparametric tests were conducted to assess significant differences in minutiae counts and image quality scores based on the force level. The results reveal a significant difference in image quality score based on the force level and each sensor technology, yet there is no significant difference in minutiae count based on the force levels of the capacitance sensor. The image quality score, shown to be effected by force and sensor type, is one of many factors that influence the system matching performance, yet the removal of low quality images does not improve the system performance at each force level.

M. R. Girgisa et al., [7] proposed a method to describe a fingerprint matching based on lines extraction and graph matching principles by adopting a hybrid scheme which consists of a genetic algorithm phase and a local search phase. Experimental results demonstrate the robustness of algorithm. Luping Ji, and Zhang Yi [8] proposed a method for estimating four direction orientation field by considering four steps, i) preprocessing fingerprint image, ii) determining the primary ridge of fingerprint block using neuron pulse coupled neural network, iii) estimating block direction by projective distance variance of a ridge, instead of a full block, iv) correcting the estimated orientation field. Duoqian Maio et al., [9] used principal graph algorithm by kegl to obtain principal curves for auto fingerprint identification system. From principal curves, minutiae extraction algorithm is used to extract the minutiae of the fingerprint. The experimental results shows curves obtained from graph algorithm are smoother than the thinning algorithm. Alessandra Lumini, and Loris Nanni [10] developed a method for minutiae based fingerprint and its approach to the problem as two - class pattern recognition. The obtained feature vector by minutiae matching is classified into genuine or imposter by Support Vector Machine resulting remarkable performance improvement Xifeng Tong et al., [11] proposed a method to overcome non linear distortion using Local Relative Error Descriptor (LRLED).The algorithm consists of three steps i) a pair wise alignment method to achieve fingerprint alignment ii) a matched minutiae pair set is obtained with a threshold to reduce non-matches finally iii) the LRLED – based similarity measure. LRLED is good at distinguishing between corresponding and non corresponding minutiae-pairs and works well for fingerprint minutiae matching. L. Lam et al., [12] presented a method, thinning is the process of reducing thickness of each line of patterns to just a single pixel width. The requirements of a good algorithm with respect to a fingerprint are i) the thinned fingerprint image obtained should be of single pixel width with no discontinuities ii) Each ridge should be thinned to its central pixel iii) Noise and singular pixels should be eliminated iv) no further removal of pixels should be possible after completion of thinning process. Mohamed et al., [13] presented fingerprint classification system using Fuzzy Neural Network. The fingerprint features such as singular points, positions and direction of core and delta obtained from a binarised fingerprint image. The method is producing good classification results. Ching-Tang Hsieh and Chia-Shing – Hu [14] has developed anoid method for Fingerprint recognition. Ridge bifurcations are used as minutiae and ridge bifurcation algorithm with excluding the noise–like points are proposed. Experimental results show the humanoid fingerprint recognition is robust, reliable and rapid.

Lie Wei [15] proposed a method for rapid singularities searching algorithm which uses delta field Poincare index and a rapid classification algorithm to classify the fingerprint in to 5 classes. The detection algorithm searches the direction field which has the larger direction changes to get the singularities. Singularities detection is used to increase the accuracy. Hartwig Fronthaler, et al., [16] Proposed fingerprint enhancement to improve the matching performance and computational efficiency by using an image scale pyramid and directional filtering in the spatial domain. Mana Tarjoman and Shaghayegh Zarei [17] introduced structural approach to fingerprint classifications by using the directional image of fingerprint instead of singularities. Directional image includes dominant direction of ridge lines. Bhupesh Gour et al., [18] have developed a method for extraction of minutiae from fingerprint images using midpoint ridge contour representation. The first step is segmentation to separate foreground from background of fingerprint image. A 64 x 64 region is extracted from fingerprint image. The grayscale intensities in 64 x 64 regions are normalized to a constant mean and variance to remove the effects of sensor noise and grayscale variations due to finger pressure differences. After the normalization the contrast of the ridges are enhanced by filtering 64 x 64 normalized windows by appropriately tuned Gabor filter. Processed fingerprint image is then scanned from top to bottom and left to right and transitions from white (background) to black (foreground) are detected. The length vector is calculated in all the eight directions of contour. Each contour element represents a pixel on the contour, contains fields for the x, y coordinates of the pixel. The proposed method takes less and do not detect any false minutiae. Sharath Pankanti et al., [19] proposed Scale Invariant Feature Transformation (SIFT) to represent and match the fingerprint. By extracting characteristic SIFT feature points in scale space and perform matching based on the texture information around the feature points. The combination of SIFT and conventional minutiae based system achieves significantly better performance than either of the individual schemes. Manvjeet Kaur et al., [20] have introduced combined methods to build a minutia extractor and a minutia matcher. Segmentation with Morphological operations used to improve thinning, false minutiae removal, minutia marking. Haiping Lu et al., [21] proposed an effective and efficient algorithm for minutiae extraction to improve the overall performance of an automatic fingerprint identification system because it is very important to preserve true minutiae while removing spurious minutiae in post-processing. The proposed novel fingerprint image post-processing algorithm makes an efforts to reliably differentiate spurious minutiae from true ones by making use of ridge number information, referring to original gray-level image, designing and arranging various processing techniques properly, and also selecting various processing parameters carefully. The proposed post-processing algorithm is effective and efficient. Prabhakar S, Jain. A.K. et al., [22] has developed filter-based representation technique for fingerprint identification. The technique exploits both local and global characteristics in a fingerprint to make identification. Each fingerprint image is filtered in a number of directions and a 640-dimensinal feature vector is extracted in the central region of the fingerprint. The feature vector is compact and requires only 640 bytes. The matching stage computes the Euclidian distance between the template finger code and the input finger code. The method gives good matching with high accuracy. Ballan M [23] introduced Directional Fingerprint Processing using fingerprint smoothing, classification and identification based on the singular points (delta and core points) obtained from the directional histograms of a fingerprint. Fingerprints are classified into two main categories that are called Lasso and Wirbel. The process includes directional image formation, directional image block representation, singular point detection and decision. The method gives matching decision vectors with minimum errors, and method is simple and fast.

### III. MODEL

In this section the definitions and FRMSM model are discussed

A. Definitions:

*Termination* : The location where a ridge comes to an end.
*Bifurcation* : The location where a ridge divides into two separate ridges.
*Binarization* : The process of converting the original grayscale image to a black-and white image.
*Thinning* : The process of reducing the width of each ridge to one pixel
*Termination Angle* : The angle between the horizontal and the direction of the ridge.
*Bifurcation Angle* : The angle between the horizontal and the direction of the valley ending between the bifurcations.
*False Matching Ratio* : It is the probability that the system will decide to allow access to an *(FMR)* imposter is given in an equation (1)

$$FMR = \frac{FalseMatches}{\text{Im} posterAttempts} \quad \text{-----------------} \quad (1)$$

The imposter attempts are implemented by matching each input image with all the template images. False match was recorded for each imposter attempt when the matching score was greater than the established threshold.

(viii) *False Non Matching Ratio (FNMR):* It is the probability that the system denies access to an approved user is given in an equation (2)

$$FNMR = \frac{FalseNonMatches}{EnrolleAttempts} \quad \text{---------------} \quad (2)$$

Enrollee attempts are implemented by matching each input image with corresponding template image, hence it is one-to-one matching. A False Non-match was recorded when the matching score between an enrolee and its template was less than the established threshold.

*(ix) Matching Score*: it is used to calculate the matching score between the input and template data is given in an equation (3)

$$Matching\, score = \frac{Matching\, Minutiae}{Max(NT, NI)} \quad \text{------ (3)}$$

Where, *NT* and *NI* represent the total number of minutiae in the template and input matrices respectively. By this definition, the matching score takes on a value between 0 and 1. Matching score of 1 and 0 indicates that data matches perfectly and data is completely mismatched respectively.

**B. Model**

Figure 1 gives the block diagram of FRMSM which is used to match the test fingerprint with the template database using Minutia Matching Score.

Fingerprint Image: The input fingerprint image is the gray scale image of a person, which has intensity values ranging from 0 to 255. In a fingerprint image, the ridges appear as dark lines while the valleys are the light areas between the ridges. Minutiae points are the locations where a ridge becomes discontinuous. A ridge can either come to an end, which is called as termination or it can split into two ridges, which is called as bifurcation. The two minutiae types of terminations and bifurcations are of more interest for further processes compared to other features of a fingerprint image.

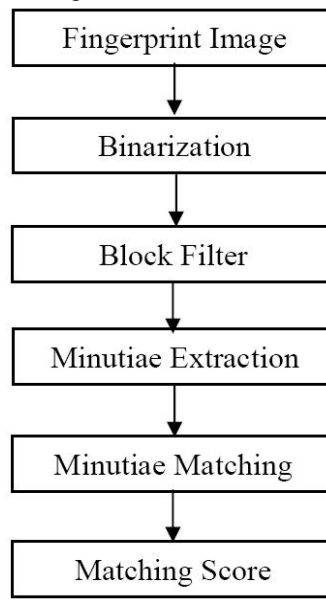

**Fig 1: Block Diagram of FRMSM.**

**Binarization:** The pre-processing of FRMSM uses Binarization to convert gray scale image into binary image by fixing the threshold value. The pixel values above and below the threshold are set to '1' and '0' respectively. An original image and the image after Binarization are shown in the Figure 2.

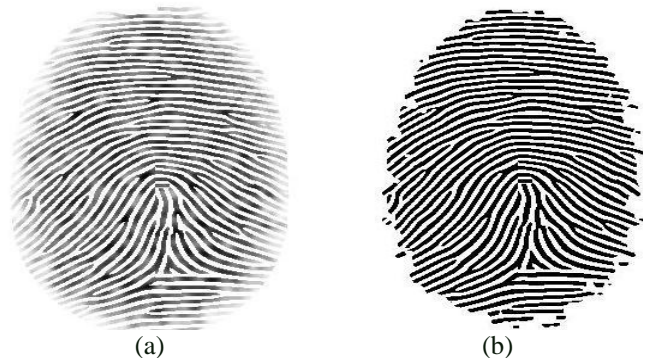

(a)          (b)
**Fig 2: (a) Original Fingerprint  (b) Binarized image.**

**Block Filter:** The binarized image is thinned using Block Filter to reduce the thickness of all ridge lines to a single pixel width to extract minutiae points effectively. Thinning does not change the location and orientation of minutiae points compared to original fingerprint which ensures accurate estimation of minutiae points. Thinning preserves outermost pixels by placing white pixels at the boundary of the image, as a result first five and last five rows, first five and last five columns are assigned value of one. Dilation and erosion are used to thin the ridges. A binarized Fingerprint and the image after thinning are shown in Figure 3.

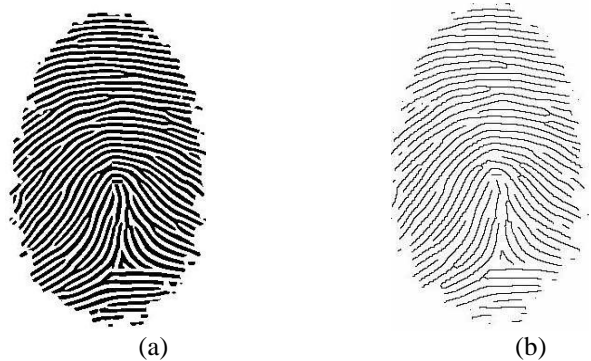

(a)  (b)

**Fig 3: (a) Binarized Fingerprint   (b) Image after thinning**

**Minutiae Extraction:** The minutiae location and the minutiae angles are derived after minutiae extraction. The terminations which lie at the outer boundaries are not considered as minutiae points, and Crossing Number is used to locate the minutiae points in fingerprint image. Crossing Number is defined as half of the sum of differences between intensity values of two adjacent pixels. If crossing Number is 1, 2 and 3 or greater than 3 then minutiae points are classified as Termination, Normal ridge and Bifurcation respectively, is shown in figure 4.

| | |
|---|---|
| ▦ | Crossing Number =2. Normal ridge pixel. |
| ▦ | Crossing Number =1. Termination point. |
| ▦ | Crossing Number =3. Bifurcation point. |

**Fig 4: Crossing Number and Type of Minutiae.**

To calculate the bifurcation angle, we use the advantage of the fact that termination and bifurcation are dual in nature. The termination in an image corresponds to the bifurcation in its negative image hence by applying the same set of rules to the negative image, we get the bifurcation angles.
Figure 5 shows the original image and the extracted minutiae points. Square shape shows the position of termination and diamond shape shows the position of bifurcation as in figure 5 (b)

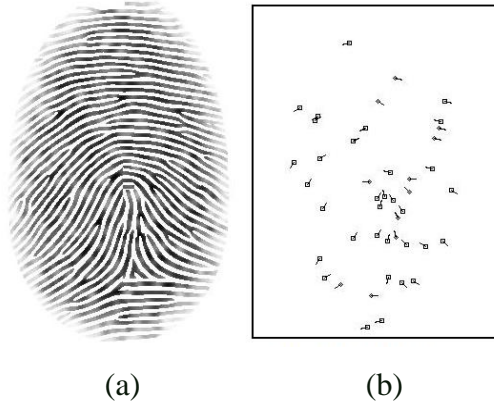

(a)            (b)
**Fig 5: (a) Gray-scale Fingerprint (b) Minutiae points.**

**Minutiae Matching:** To compare the input fingerprint data with the template data Minutiae matching is used. For efficient matching process, the extracted data is stored in the matrix format. The data matrix is as follows.
Number of rows: Number of minutiae points.
Number of columns: 4
Column 1: Row index of each minutia point.
Column 2: Column index of each minutia point.
Column 3: Orientation angle of each minutia point.
Column 4: Type of minutia. (A value of '1' is assigned for termination, and '3' is assigned for bifurcation).
During the matching process, each input minutiae point is compared with template minutiae point. In each case, template and input minutiae are selected as reference points for their respective data sets. The reference points are used to convert the remaining data points to polar coordinates. The Equation (4) is used to convert the template minutiae from row and column indices to polar coordinates

$$\begin{pmatrix} r_k^T \\ \phi_k^T \\ \theta_k^T \end{pmatrix} = \begin{pmatrix} \sqrt{(row_k^T - row_{ref}^T)^2 + (col_k^T - col_{ref}^T)^2} \\ \tan^{-1}\left(\dfrac{row_k^T - row_{ref}^T}{col_k^T - col_{ref}^T}\right) \\ \theta_k^T - \theta_{ref}^T \end{pmatrix}$$

--------------- (4)

Where, for a template image,

$r_k^T$ = radial distance of $k^{th}$ minutiae.

$\phi_k^T$ = radial angle of $k^{th}$ minutiae.

$\theta_k^T$ = orientation angle of $k^{th}$ minutiae.

$row_{ref}^T, col_{ref}^T$ = row index and column index of reference points currently being considered.

Similarly the input matrix data points are converted to polar coordinates using the Equation (5)

$$\begin{pmatrix} r_m^I \\ \phi_m^I \\ \vartheta_m^I \end{pmatrix} = \begin{pmatrix} \sqrt{(row_m^I - row_{ref}^I)^2 + (col_m^I - col_{ref}^I)^2} \\ \tan^{-1}\left(\dfrac{row_m^I - row_{ref}^I}{col_m^I - col_{ref}^I}\right) + rotatevalue(k,m) \\ \theta_m^I - \theta_{ref}^I \end{pmatrix}$$

-------------- (5)

*Rotate values (k, m)* represents the difference between the orientation angles of *Tk* and *Im*. *Tk* and *Im* represent the extracted data in all the columns of row *k* and row *m* in the template and input matrices, respectively.

### IV. ALGORITHM

Problem definition**:** Given the test Fingerprint Image the objectives are,
1. Pre-processing the test Fingerprint.
2. Extract the minutiae points.
3. Matching test Fingerprint with the database.

Table 1 gives the algorithm for fingerprint verification, in which input test fingerprint image is compared with template fingerprint image, for recognition.

**Table 1: Algorithm of FRMSM**

| |
|---|
| Input: Gray-scale Fingerprint image. |
| Output: Verified fingerprint image with matching score. |
| 1. Fingerprint is binarized |
| 2. Thinning on binarized image |
| 3. Minutiae points are extracted. Data matrix is generated to get the position, orientation and type of minutiae. |
| 4. Matching of test fingerprint with template |
| 5. Matching score of two images is computed, if matching score is 1 images are matched and if it is 0 then they are mismatched. |

### V. PERFORMANCE ANALYSIS AND RESULTS

For performance analysis, we considered large fingerprint database images having different patterns such as fingerprint left loop, right loop, whorl and arch as shown in the Figure 6.

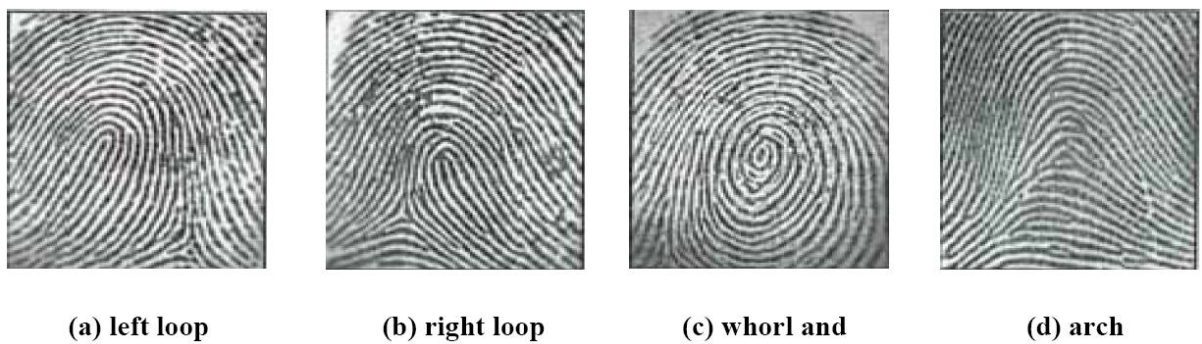

(a) left loop      (b) right loop      (c) whorl and      (d) arch

**Fig 6: Samples of fingerprint images**

Table 2 gives the comparison of False Non Matching Ratio (FNMR) and False Matching Ratios (FMR) for existing method of Fingerprint Recognition Fuzzy Neural Network (FRFNN) and proposed method of Fingerprint Recognition using Minutia Score Matching method (FRMSM). It is observed that the False Non Matching Ratio for both the methods is zero and False Matching Ratio for existing method is 0.23 whereas for the proposed method FRMSM is 0.026.

**Table 2: Comparison of FNMR and FMR for FRFNN and FRMSM.**

|  | FRFNN | FRMSM |
|---|---|---|
| **FNMR** | 0.00 | 0.00 |
| **FMR** | 0.23 | 0.026 |

### VI. CONCLUSION

In this paper, we presented Fingerprint matching using FRMSM. The pre-processing the original fingerprint involves image binarization, ridge thinning, and noise removal. Fingerprint Recognition using Minutia Score Matching method is used for matching the minutia points. The proposed method FRMSM gives better FMR values compared to the existing method.